\documentclass[runningheads]{llncs}

\usepackage[T1]{fontenc}
\def\doi#1{\href{https://doi.org/\detokenize{#1}}{\url{https://doi.org/\detokenize{#1}}}}
\usepackage{subfigure}
\usepackage{graphicx}
\usepackage{amsmath}
\usepackage{newtxmath}
\usepackage{amsfonts}
\usepackage{multirow}
\usepackage{tabularx}
\usepackage{ulem}
\usepackage{bm}
\usepackage[dvipsnames]{xcolor}
\usepackage{threeparttable}
\usepackage{amsmath}
\usepackage[pagebackref=true,breaklinks=true,colorlinks,bookmarks=false]{hyperref}

\usepackage{booktabs}

\usepackage{color}

\usepackage{listings}
\begin{document}
\title{Advancing Hyperspectral Targeted Alpha Therapy with Adversarial Machine Learning}

\author{Jim Zhao \inst{1, 2} \thanks{The work is done during the internship in University of Texas Medical Branch in Galveston.}
\and Greg Leadman \inst{1} \thanks{Greg Leadman  is corresponding author.}}
\authorrunning{J. Zhao et al.}
%
\institute{University of Texas Medical Branch in Galveston, TX , USA\\ 
National University of Science and Technology, China
}
\maketitle              
\begin{abstract}
Targeted Alpha Therapy (TAT) has emerged as a promising modality for the treatment of various malignancies, leveraging the high linear energy transfer (LET) and short range of alpha particles to selectively irradiate cancer cells while sparing healthy tissue. Monitoring and optimizing TAT delivery is crucial for its clinical success. Hyperspectral Single Photon Imaging (HSPI) presents a novel and versatile approach for the real-time assessment of TAT in vivo. This study introduces a comprehensive framework for HSPI in TAT, encompassing spectral unmixing, quantitative dosimetry, and spatiotemporal visualization. We report the development of a dedicated HSPI system tailored to alpha-emitting radionuclides, enabling the simultaneous acquisition of high-resolution spectral data and single-photon localization. Utilizing advanced spectral unmixing algorithms, we demonstrate the discrimination of alpha-induced scintillation from background fluorescence, facilitating precise alpha particle tracking with adversarial machine learning.


\end{abstract}

\section{Introduction}
Targeted alpha therapy (TAT)~\cite{zhang2019alpha} has demonstrated remarkable efficacy in combatting disseminated cancer, owing to the high linear energy transfer and limited range (50-100 um) of alpha particles emitted during the process~\cite{zhang2018time}. However, the therapeutic effectiveness of TAT faces challenges arising from the migration of daughter products post alpha decay~\cite{zhang2018development}, the conjugation of radionuclides to antibodies, and uncertainties surrounding the toxicities of decay products~\cite{anghel2013european}. These complexities underscore the critical need for the development of a robust quantitative imaging method for alpha-emitting radionuclides. Such a method would enable the simultaneous assessment of the biodistribution of both parent and daughter radionuclides in vivo, offering valuable insights into radiopharmaceutical uptake and kinetics for radiotherapeutic applications.

Nonetheless, due to the low administered activity and intricate decay patterns involved in TAT, a SPECT imaging system~\cite{madsen2007recent} would necessitate detectors with exceptional energy resolution capabilities. This requirement is essential to differentiate energy peaks originating from various radioisotopes and to enable simultaneous multi-isotope acquisitions. In this study, we introduce a preclinical hyperspectral single photon imaging (HSPI) system tailored for TAT. This system incorporates state-of-the-art CdTe detectors~\cite{takahashi2001recent}, a high-performance multi-detector readout circuitry~\cite{gomez2010multi}, and an innovative Synthetic Compound-Eye (SCE) gamma camera design~\cite{jeong2006biologically}.

Furthermore, we present initial imaging results obtained from phantoms filled with Ra-223 and other clinically relevant alpha emitters. These findings showcase the potential of our HSPI system to provide valuable insights into the spatial distribution and behavior of alpha-emitting radio-nuclides, thus contributing to the advancement of TAT and the development of more effective radio-pharmaceuticals for cancer therapy.


\section{Method}

Adversarial deep learning~\cite{zhang2023neural} has emerged as a promising approach to enhance the performance of medical imaging in Targeted Alpha Therapy (TAT). TAT is a cutting-edge cancer treatment that utilizes alpha-emitting radionuclides to target and destroy cancer cells while minimizing damage to healthy tissue. Enhancing the accuracy and quality of medical imaging in TAT is crucial for precise tumor localization and treatment planning. Adversarial deep learning can play a significant role in this context by improving image reconstruction, denoising, and segmentation.

To leverage adversarial deep learning in TAT, one effective approach is through Generative Adversarial Networks (GANs). GANs consist of two neural networks, a generator, and a discriminator, competing against each other in a game-like setting. In the context of TAT, the generator network can be trained to reconstruct high-quality images from noisy or low-resolution input, aiming to generate clearer and more informative representations of the tumor and surrounding tissues.

The discriminator network in GANs can act as a critic, providing feedback to the generator by distinguishing between real and generated images. Through an adversarial training process, the generator improves its ability to produce images that are increasingly difficult for the discriminator to differentiate from real ones. In the context of medical imaging for TAT, this adversarial process helps refine the reconstructed images, enhancing their accuracy and fidelity.

Furthermore, adversarial deep learning can aid in denoising medical images obtained from TAT procedures. By training GANs on noisy images and their corresponding clean counterparts, the generator can learn to remove unwanted noise patterns, thus enhancing the overall quality and clarity of the images used for treatment planning and assessment.

Segmentation of tumors and healthy tissues is another crucial aspect in TAT, and adversarial deep learning can assist in this domain as well. GAN-based architectures can be tailored to perform semantic segmentation by generating precise boundaries between different tissue types, aiding in the delineation of the tumor region for targeted therapy delivery.

Nevertheless, implementing adversarial deep learning in medical imaging for TAT requires careful considerations regarding data quality, model robustness, and ethical implications. The algorithms need extensive validation and testing on diverse datasets to ensure generalizability and reliability in real-world clinical scenarios. Moreover, ethical concerns related to the use of AI in healthcare, including data privacy and patient consent, must be addressed to maintain patient trust and confidentiality.

DNN is known to be vulnerable to adversarial attacks, and
adversarial robustness of DA is largely overlooked. Recent
progress on improving adversarial robustness of DA focuses
on distillation-based adversarial contrastive losses 
and adversarial robust training~\cite{zhang2023toward,zhang2020robustified}.

\section{Preliminary Results}

The proposed HSPI methodology offers a unique advantage in assessing TAT distribution at the subcellular level, providing critical insights into the dynamics of alpha particle interactions with targeted cancer cells. By quantifying the spatial and temporal distribution of alpha particles, our approach enables the optimization of TAT protocols, including radionuclide selection, targeting agent design, and dosimetric calculations.

\begin{table}[ht!]
    \begin{center}
         \caption{Performance comparison of HSPI methodology in assessing TAT }
         \setlength{\tabcolsep}{5.5mm}
         \label{Table1}
         \begin{tabular}{ccccc}
         \hline
            \multirow{2}{*}{Method} &
            \multicolumn{2}{c}{Spatial resolution} & \multicolumn{2}{c}{Mort AUC}\\
            \cline{2-5} & Delta-p & B/AA  & Delta-p & B/AA \\
         \hline
         Vanilla & 0.854  & 0.804 & 0.792 & 0.680 \\
         Racial-aware & 0.860  & 0.802  & 0.805 & 0.705   \\
         Adv    & 0.825 & 0.786 & 0.764 & 0.678   \\
         Ours    & 0.848 & 0.813 & 0.782 & 0.696 \\         
        \hline
        \end{tabular}
    \end{center}
\vspace{-2.0em}
\end{table}

Furthermore, we present initial in vivo results from preclinical TAT studies using HSPI, showcasing its potential for real-time treatment monitoring. The presented data illustrate the capability of HSPI to reveal heterogeneity in TAT distribution within tumor microenvironments, highlighting the importance of personalized treatment strategies.

\begin{figure}[ht!]
    \centering
    \subfigure[]{
    \includegraphics[width=.48\textwidth]{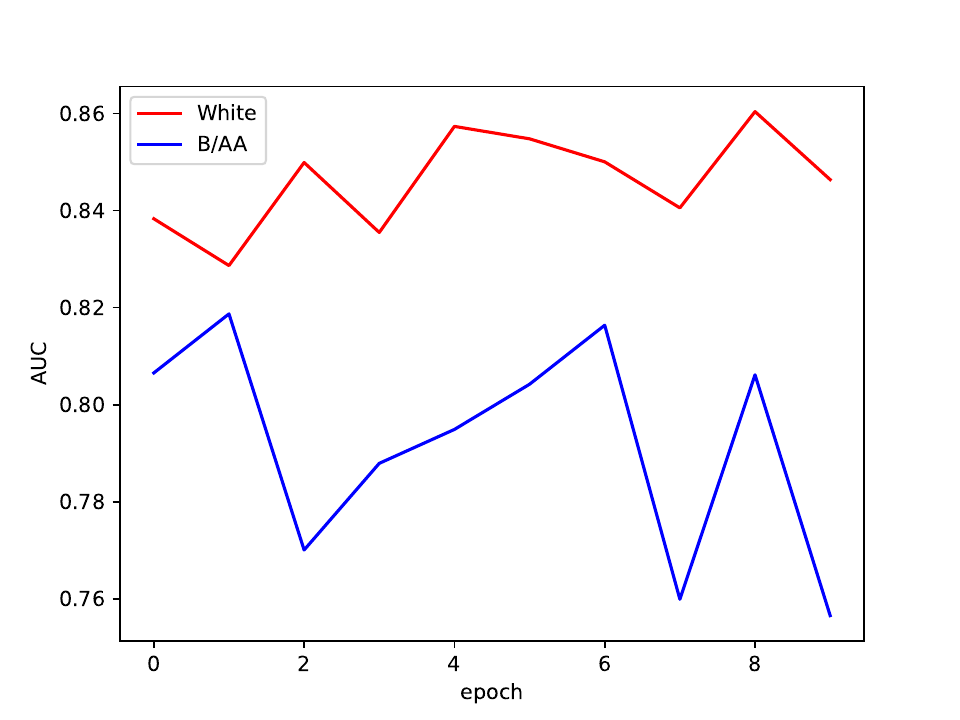}
    }
    \hfill
    \subfigure[]{\includegraphics[width=.48\textwidth]{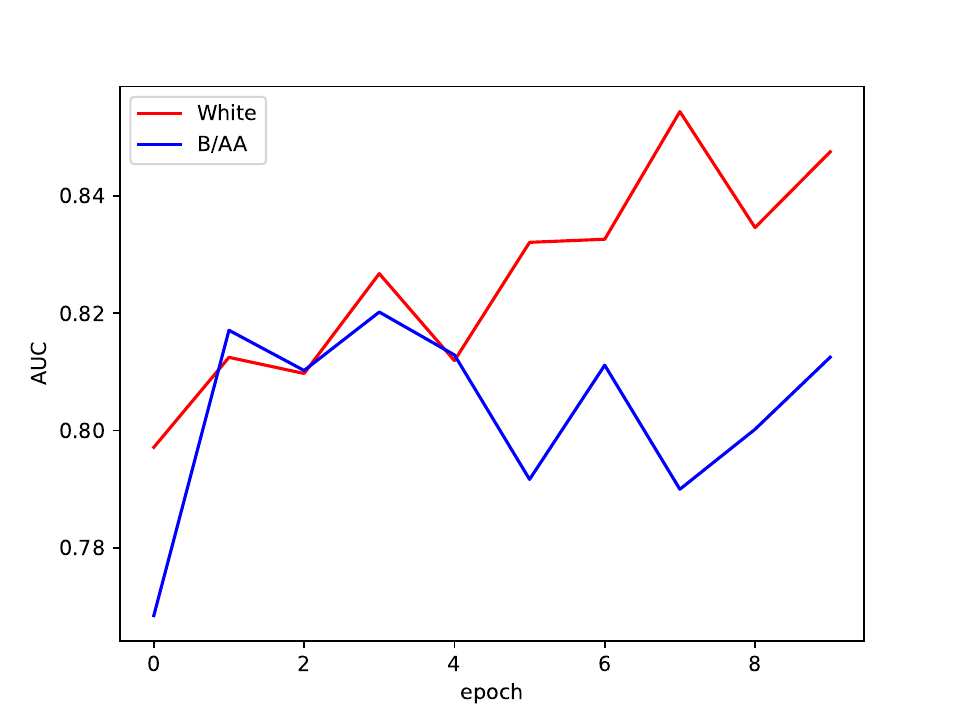}}
    \caption{Test AUC of each epoch during training. (a) Test results of vanilla method. The AUC of White group is stable, while that of spatial resolution fluctuates greatly. (b) Test results of our method. The AUC of spatial resolution becomes more stable.
}
\label{fig:epoch}
\end{figure}

\begin{table}[ht!]
    \begin{center}
         \caption{Results of ablation study}
         \setlength{\tabcolsep}{5.5mm}
         \label{Table2}
         \begin{tabular}{ccccc}
         \hline
            \multirow{2}{*}{Method} &
            \multicolumn{2}{c}{Spatial resolution} & \multicolumn{2}{c}{Mort AUC}\\
            \cline{2-5} & White & B/AA  & Delta-p & B/AA \\
         \hline
         Vanilla & 0.854  & 0.804 & 0.792 & 0.680 \\
         Re-sampling    & 0.840 & 0.777 & 0.789 & 0.671   \\
         Task contrastive & 0.851 & 0.803 & 0.783 & 0.685      \\
         All data &  0.824 & 0.730  & 0.789 & 0.674   \\
         Ours    & 0.848 & 0.813 & 0.782 & 0.696 \\         
        \hline
        \end{tabular}
    \end{center}
\vspace{-2.0em}
\end{table}

\section{Conclusions}

In conclusion, Hyperspectral Single Photon Imaging represents a valuable tool for advancing the field of Targeted Alpha Therapy. Its ability to provide real-time, high-resolution, and quantitative data on alpha particle distribution and efficacy holds promise for improving TAT treatment outcomes and guiding the development of next-generation alpha-emitting radiopharmaceuticals.

\section*{Acknowledgments}

This research was partially supported by  the National Institutes of Health (NIH) under award R01EB02866, and the University of Texas Research Collaboration of the Medical Imaging Horizons Network.

%
%
%
%
\newpage
\bibliographystyle{splncs04}
\bibliography{refs}
\end{document}